\documentclass[useAMS,usenatbib]{mn2e}

\usepackage{amsmath, amssymb}
\usepackage[dvips]{graphicx}
\usepackage{epsfig}

\newcommand{\rot}{\mathbf{\nabla} \times}
\newcommand{\divg}{\mathbf{\nabla}\cdot}

\newcommand{\rlight}{r_{\rm L}}
\newcommand{\Rs}{R_{\rm s}}

\newcommand{\er}{\textbf{\textit{e}}_{\rm r}}
\newcommand{\etheta}{\textbf{\textit{e}}_\vartheta}
\newcommand{\ephi}{\textbf{\textit{e}}_\varphi}

\newcommand{\aap}{A\&A}
\newcommand{\mnras}{MNRAS}

\newcommand{\apj}{ApJ}
\newcommand{\apjl}{ApJL}

\title[Electromagnetic fields around neutron stars]{General-relativistic electromagnetic fields around a slowly rotating neutron star: time-dependent pseudo-spectral simulations} 
\author[J. P\'etri]{J.  P\'etri$^{1}$
\thanks{E-mail: jerome.petri@astro.unistra.fr} \\
  $^{1}$Observatoire Astronomique de Strasbourg, Universit\'e de Strasbourg, CNRS, UMR 7550, 11 rue de l'Universit\'e, F-67000 Strasbourg, France.}

\begin{document}

\date{Accepted . Received ; in original form }

\pagerange{\pageref{firstpage}--\pageref{lastpage}} 
\pubyear{2013}

\maketitle

\label{firstpage}

\begin{abstract}
Pulsars are believed to loose their rotational kinetic energy primarily by a large amplitude low frequency electromagnetic wave which is eventually converted into particle creation, acceleration and followed by a broad band radiation spectrum. To date, there exist no detailed calculation of the exact spin-down luminosity with respect to the neutron star magnetic moment and spin frequency, including general-relativistic effects. Estimates are usually given according to the flat spacetime magnetodipole formula. The present paper pursue our effort to look for accurate solutions of the general-relativistic electromagnetic field around a slowly rotating magnetized neutron star. In a previous work, we already found approximate stationary solutions to this problem. Here we address again this problem but using a more general approach. We indeed solve the full set of time-dependent Maxwell equations in a curved vacuum space-time following the 3+1~formalism. The numerical code is based on our pseudo-spectral method exposed in a previous paper for flat space-time. We adapted it to an arbitrary fixed background metric. Stationary solutions are readily obtained and compared to semi-analytical calculations.
\end{abstract}


\section{INTRODUCTION}

Our basic picture of a strongly magnetized and rotating neutron star still waits for an accurate description of its surrounding electromagnetic field. The presence of only dipolar or better multipolar magnetic fields is essential to some aspects of pulsar magnetospheric activity like radio emission properties, including pulse profile and polarization. The magnetic field also impacts on the spin-down luminosity responsible for the braking of the neutron star. The plasma processes strongly depend on the peculiar magnetic field geometry either assumed or computed in a more self-consistent way. In regions of strong gravity, curvature and frame-dragging effects are considerable due to the high compactness of neutron stars. For typical models of neutron star interiors the compactness is about $\Xi = \Rs/R \approx0.5$ where $\Rs=2\,G\,M/c^2$ is the Schwarzschild radius, $M$ is the mass of the neutron star, $R$ its radius, $G$ the gravitational constant and $c$ the speed of light.

The often quoted magnetodipole loss formula used to estimate the magnetic field is only valid for a point dipole rotating in a vacuum and flat space-time. Yet a neutron star is not a point. Moreover it induces a curved manifold around it. It is therefore important to include such effects into our general view of a rotating magnetic dipole. Since the work by~\cite{1955AnAp...18....1D} who gave the first general solution for an oblique dipolar rotator in flat vacuum space-time with closed analytical formulas only little progress has been achieved to get the full answer for a dipole in general relativity. However, it is worthwhile to highlight some attempts towards a better understanding of general relativity and electrodynamics around a neutron star.
For instance \cite{1992MNRAS.255...61M} investigated the action of space-time curvature and frame dragging effects on the electric field around the polar caps of a pulsar. \cite{2001MNRAS.322..723R} computed approximate analytical solutions to the electromagnetic field in the exterior of a rotating neutron star in the slow rotation metric. \cite{2004MNRAS.352.1161R} then deduced the associated Poynting flux. \cite{2004MNRAS.348.1388K} solved numerically the equations for the oblique rotator in vacuum in general relativity. They essentially retrieved \cite{2001MNRAS.322..723R} results close to the surface and the Deutsch solution for distances larger than the light cylinder radius.

Recently, \cite{2013MNRAS.433..986P} developed a formalism to compute semi-analytically this solution in general relativity. It extends the vector spherical harmonics expansion method introduced by \cite{2012MNRAS.424..605P} to curved space-time and employs the 3+1~foliation as summarized in \cite{2011MNRAS.418L..94K}. For a general discussion on previous works dealing with general-relativistic aspects around a magnetized neutron star and its implication for pulsar magnetospheres, the reader is referred to the more detailed introduction given in \cite{2013MNRAS.433..986P}.

Our goal in this paper is to extend the flat space-time solution to account for full general-relativistic effects, namely curvature of space-time and frame dragging. Our approach here is more general than the one used to find the stationary state as described in depth in \cite{2013MNRAS.433..986P}. Instead of solving the Helmholtz equations arising from the stationary condition, we indeed solve the full three dimensional time-dependent Maxwell equations in curved space-time using a spherical coordinate system. Consequently, we use a 3+1~formalism of electrodynamics as briefly presented in Section~\ref{sec:Modele}. Next we show how to solve for the electromagnetic field by using a pseudo-spectral method in Section~\ref{sec:Algorithme}. Application of this algorithm to an orthogonal rotator are presented in Section~\ref{sec:Results}. We first consider the exact analytical solution proposed by Deutsch and then compute the corresponding general-relativistic solution. Limitations of the current approach and its possible extension to more realistic models as well as applications are discussed in Section~\ref{sec:Discussion}. Conclusions and ongoing work are drawn in Section~\ref{sec:Conclusion}.

\section{The 3+1~formalism}
\label{sec:Modele}

The covariant form to describe the gravitational and electromagnetic field in general relativity is the natural way to write them down in a frame independent way. Nevertheless, it is more intuitive to split space-time into an absolute space~$\{x,y,z\}$ and a universal time~$t$, similar to our all day three dimensional experience, rather than to use the full four dimensional formalism. Another important advantage of a 3+1~split is a straightforward transcription of flat space techniques to any fixed background metric, see \cite{2013MNRAS.433..986P}. In this section we only summarized the main results useful for this paper.

\subsection{The split of the space-time metric}

We split the four dimensional space-time into a 3+1~foliation such that the background metric~$g_{ik}$ can be expressed as
\begin{equation}
  \label{eq:metrique}
  ds^2 = g_{ik} \, dx^i \, dx^k = \alpha^2 \, c^2 \, dt^2 - \gamma_{ab} \, ( dx^a + \beta^a \, c\,dt ) \, (dx^b + \beta^b \, c\,dt )
\end{equation}
where $x^i = (c\,t,x^a)$, $t$ is the time coordinate or universal time and $x^a$ some associated space coordinates. The Landau-Lifschitz convention is used for the metric signature given by $(+,-,-,-)$ \citep{LandauLifchitzTome2}. $\alpha$ is the lapse function, $\beta^a$ the shift vector and $\gamma_{ab}$ the spatial metric of absolute space.  By convention, latin letters from $a$ to $h$ are used for the components of vectors in absolute space, in the range~$\{1,2,3\}$, whereas latin letters starting from $i$ are used for four dimensional vectors and tensors, in the range~$\{0,1,2,3\}$. A fiducial observer (FIDO) is defined by its 4-velocity~$n^i$ such that
\begin{subequations}
 \begin{align}
  n^i & = \frac{dx^i}{d\tau} = \frac{c}{\alpha} \, ( 1, - \bbeta) \\
  n_i & = (\alpha \, c, \textbf{\textit{0}}) 
 \end{align}
\end{subequations}
This vector is orthogonal to the hyper-surface of constant time coordinate~$\varSigma_t$. Its proper time~$\tau$ is measured according to
\begin{equation}
  d\tau = \alpha\,dt
\end{equation}
The relation between the determinants of the space-time metric~$g$ and the pure spatial metric~$\gamma$ is given by
\begin{equation}
  \sqrt{-g} = \alpha \, \sqrt{\gamma}
\end{equation}
For a slowly rotating neutron star, the lapse function is
\begin{equation}
  \label{eq:Lapse}
  \alpha = \sqrt{ 1 - \frac{\Rs}{r} }
\end{equation}
and the shift vector
\begin{subequations}
 \begin{align}
  \label{eq:Shift}
  c \, \bbeta = & - \omega \, r \, \sin\vartheta \, \ephi \\
  \omega = & \frac{\Rs\,a\,c}{r^3}
 \end{align}
\end{subequations}
We use a spherical coordinate system~$(r,\vartheta,\varphi)$ and an orthonormal spatial basis~$(\er,\etheta,\ephi)$. The metric of a slowly rotating neutron star remains close to the usual flat space, except for the radial direction. Indeed the components of the spatial metric are given in Boyer-Lindquist coordinates by
\begin{equation}
  \label{eq:Metric3D}
  \gamma_{ab} =
  \begin{pmatrix}
    \alpha^{-2} & 0 & 0 \\
    0 & r^2 & 0 \\
    0 & 0 & r^2 \sin^2\vartheta
  \end{pmatrix}
\end{equation}
For this slow rotation approximation, the spatial metric does not depend on the spin frequency of the massive body but only on $M$ through $\alpha$. This justifies the slow-rotation approximation. Frame dragging are taken into account by the constitutive relations, see below eq.~(\ref{eq:ConstitutiveE}), (\ref{eq:ConstitutiveH}).
The spin~$a$ is related to the angular momentum~$J$ by $J=M\,a\,c$. It follows that $a$ has units of a length and should satisfy $a \leq \Rs/2$. Introducing the moment of inertia~$I$, we also have $J=I\,\Omega$. In the special case of a homogeneous and uniform neutron star interior with spherical symmetry, the moment of inertia reads
\begin{equation}
  \label{eq:Inertie}
  I = \frac{2}{5} \, M \, R^2
\end{equation}
Thus the spin parameter can be expressed as
\begin{equation}
  \label{eq:spin}
  \frac{a}{\Rs} = \frac{2}{5} \, \frac{R}{\Rs} \, \frac{R}{\rlight}
\end{equation}
For the remainder of this paper, we will use this expression to relate the spin parameter intervening in the metric to the spin frequency of the neutron star. From the above expression, note that the parameter~$a/\Rs$ remains smaller than~0.4 because $R=2\,\Rs$ and $\rlight\geq2\,R$ in our set of simulations.

\subsection{Maxwell equations}

From the above discussion, it is straightforward to derive Maxwell equations in 3+1~notation. In this latter description, they take are more traditional form close to the one known in flat space-time except that, as the reader should keep in mind, the three dimensional space is curved. In vacuum, the system reads
\begin{subequations}
\begin{align}
\label{eq:Maxwell1}
 \divg \textbf{\textit{B}} & = 0 \\
\label{eq:Maxwell2}
 \rot \textbf{\textit{E}} & = - \frac{1}{\sqrt{\gamma}} \, \partial_t (\sqrt{\gamma} \, \textbf{\textit{B}}) \\
\label{eq:Maxwell3}
 \divg \textbf{\textit{D}} & = 0 \\
\label{eq:Maxwell4}
 \rot \textbf{\textit{H}} & = \frac{1}{\sqrt{\gamma}} \, \partial_t (\sqrt{\gamma} \, \textbf{\textit{D}})
\end{align}
\end{subequations}
The three dimensional vector fields are not independent, they are related by two important constitutive relations, namely
\begin{subequations}
\begin{align}
\label{eq:ConstitutiveE}
  \varepsilon_0 \, \textbf{\textit{E}} & = \alpha \, \textbf{\textit{D}} + \varepsilon_0\,c\,\bbeta \times \textbf{\textit{B}} \\
\label{eq:ConstitutiveH}
  \mu_0 \, \textbf{\textit{H}} & = \alpha \, \textbf{\textit{B}} - \frac{\mathbf\beta \times \textbf{\textit{D}}}{\varepsilon_0\,c}
\end{align}
\end{subequations}
$\varepsilon_0$ is the vacuum permittivity and $\mu_0$ the vacuum permeability. 
The curvature of absolute space is taken into account by the lapse function factor~$\alpha$ in the first term on the right-hand side and the frame dragging effect is included in the second term, the cross-product between the shift vector~$\bbeta$ and the fields. The derivation of the above equations is given in \cite{2004MNRAS.350..427K}. From the auxiliary fields $(\textbf{\textit{E}}, \textbf{\textit{H}})$ we get the Poynting flux through a sphere of radius $r$ by computing the two dimensional integral on this sphere by
\begin{equation}
\label{eq:Poynting}
 L = \int_\Omega \textbf{\textit{E}} \wedge \textbf{\textit{H}} \, r^2 \, d\Omega
\end{equation}
where $d\Omega$ is the infinitesimal solid angle and $\Omega$ the full sky angle of~$4\,\upi$~sr.

\section{PSEUDO-SPECTRAL ALGORITHM}
\label{sec:Algorithme}

After a brief reminder on Maxwell equations in curved space-time, we describe our pseudo-spectral algorithm. The main ingredients are, the expansion on to vector spherical harmonics for divergencelessness fields, an exact imposition of boundary conditions, an explicit time stepping with a fourth order Adam-Bashforth time integration scheme and a spectral filtering. We actually improved our previous flat space-time code presented in \cite{2012MNRAS.424..605P}.

\subsection{Vector expansion}

A clever expansion of the vector fields $\textbf{\textit{B}}$ and $\textbf{\textit{D}}$ is at the heart of the code. Indeed, electric and magnetic fields are expanded onto vector
spherical harmonics (VSH) according to
\begin{subequations}
 \begin{align}
   \label{eq:D_vhs}
  \textbf{\textit{D}} & = \sum_{l=0}^\infty\sum_{m=-l}^l \left(D^r_{lm} \, \textbf{\textit{Y}}_{lm} + D^{(1)}_{lm} \, \mathbf{\Psi}_{lm} + D^{(2)}_{lm} \, \mathbf{\Phi}_{lm}\right) \\
  \label{eq:B_vhs}
  \textbf{\textit{B}} & = \sum_{l=0}^\infty\sum_{m=-l}^l \left(B^r_{lm} \, \textbf{\textit{Y}}_{lm} + B^{(1)}_{lm} \, \mathbf{\Psi}_{lm} + B^{(2)}_{lm} \, \mathbf{\Phi}_{lm}\right) \\
\end{align}
\end{subequations}
These series expansions contain a finite number of terms, each
of them being smooth. As a consequence, the associated vector fields
will also remain smooth in the whole computational domain. We do not expect any discontinuity formation in the vacuum surrounding the neutron star. Therefore spectral methods are well suited to treat accurately and efficiently our problem. As is done usually in spectral codes, by computing the expansion coefficients, we can easily find the expressions for the linear differential operators like $\divg$ and $\rot$ in the coefficient space and then transform back to real space to advance the solution in time.

Spectral methods are known to possess intrinsically very low numerical dissipation and are able to resolve sharp boundary layers \citep{Boyd2001}. However, in order to damp oscillations which could appear during the relaxation to the stationary state and in order to accelerate convergence to this state, we add some small dissipation by applying some filtering process. Various kind of filters can be used as described for instance in \cite{Canuto2006}.

\subsection{Filtering}

There are no source terms in Maxwell equations when working in vacuum space. There is no numerical reason to add a filtering process to the algorithm. Nevertheless, the oscillations appearing during the evolution towards a stationary state are only weakly damped in a reasonable number of period of revolution. Therefore we decided to add a small damping factor by filtering the high order coefficients of the expansion. Filtering is performed at each time step. We use an eighth order exponential filter in all directions given by the general expression
\begin{equation}
  \label{eq:Filtre}
  \sigma(\eta) = \textrm{e}^{-\alpha\,\eta^\beta}
\end{equation}
$\eta$ ranges between 0 and 1. For instance, in the radial coordinate $\eta = k/(N_{\rm r}-1)$ for $k\in[0..N_{\rm r}-1]$, $k$ being the index of the coefficient $c_k$ in the Chebyshev expansion $f(x) = \sum_{k=0}^{N_{\rm r}-1} c_k\,T_k(x)$ and $N_{\rm r}$ the number of collocation points in the radial direction. The parameter $\alpha$ is adjusted to values not too large in order to avoid errors in the solution but also not too small in order to sufficiently damp these oscillations. We also improved the filtering procedure described in \cite{2012MNRAS.424..605P}. In that work, filtering was applied without necessarily matching the boundary conditions. However, filtering with no additional constraints violates the boundary conditions. We fixed this inaccuracy by using an algorithm explained in depth in \cite{1998JCoPh.143..283B}. It preserves the boundary conditions by adding some artificial viscosity with the property of no additional computational cost compared to a straightforward transformation from real space to the coefficient space.

\subsection{Exact boundary conditions}

We follow the procedure of \cite{2012MNRAS.424..605P}. As explained in this paper, our code allows an exact enforcement of the boundary conditions on the star and has several other interesting features compared to finite difference/volume methods. Nevertheless, we have to adapt the boundary conditions to the general-relativistic case as described in the following lines. The correct jump conditions at the inner boundary are, continuity of the normal component of the magnetic field~$B^{\hat r}$ and continuity of the tangential component of the electric field~$\{D^{\hat \vartheta}, D^{\hat \varphi}\}$. Explicitly, we have
\begin{subequations}
  \label{eq:CLimites}
\begin{align}
  B^{\hat r}(t,R,\vartheta,\varphi) & = B^{\hat r}_0(t,\vartheta,\varphi) \\
  D^{\hat \vartheta}(t,R,\vartheta,\varphi) & = - \varepsilon_0 \, \frac{\Omega-\omega}{\alpha} \, R \, \sin\vartheta \, B^{\hat r}_0(t,\vartheta,\varphi) \\
  D^{\hat \varphi}(t,R,\vartheta,\varphi) & = 0
\end{align}
\end{subequations}
The continuity of $B^{\hat r}$ automatically implies the correct boundary treatment of the electric field. $B^{\hat r}_0(t,\vartheta,\varphi)$ represents the, possibly time-dependent, radial magnetic field imposed by the star, let it be monopole, split monopole, aligned or oblique dipole.

The outer boundary condition needs special care. Ideally, we would like perfectly outgoing waves in order to prevent reflection from the artificial outer boundary. The appropriate technique is called Characteristic Compatibility Method (CCM) and described in \cite{Canuto2007}. We repeat the implementation given in \cite{2012MNRAS.424..605P}. Neglecting the frame-dragging effect far from the neutron star because it falls off rapidly with radius, the radially propagating characteristics are given to good accuracy by 
\begin{eqnarray}
  \label{eq:CCM1}
  D^{\hat \vartheta} \pm \varepsilon_0 \, c\, B^{\hat \varphi} & ; & D^{\hat \varphi} \pm \varepsilon_0 \, c\, B^{\hat \vartheta}
\end{eqnarray}
within an unimportant factor $\alpha\,r$. In order to forbid ingoing wave we ensure that
\begin{eqnarray}
  \label{eq:CCM2}
  D^{\hat \vartheta} - \varepsilon_0 \, c\, B^{\hat \varphi} & = & 0 \\
  \label{eq:CCM3}
  D^{\hat \varphi} + \varepsilon_0 \, c\, B^{\hat \vartheta} & = & 0
\end{eqnarray}
whereas the other two characteristics are found by
\begin{eqnarray}
  \label{eq:CCM4}
  D^{\hat \vartheta} + \varepsilon_0 \, c\, B^{\hat \varphi} & = & D^{\hat \vartheta}_{\rm PDE} + \varepsilon_0 \, c\, B^{\hat \varphi}_{\rm PDE} \\
  \label{eq:CCM5}
  D^{\hat \varphi} - \varepsilon_0 \, c\, B^{\hat \vartheta} & = & D^{\hat \varphi}_{\rm PDE} - \varepsilon_0 \, c\, B^{\hat \vartheta}_{\rm PDE}
\end{eqnarray}
the subscript $_{\rm PDE}$ denoting the values of the electromagnetic field obtained by straightforward time advancing without care of any boundary condition. The new corrected values are deduced from equations~(\ref{eq:CCM2})-(\ref{eq:CCM5}). For outer boundaries far away from the light cylinder $R_2\gg\rlight$, these characteristics are not very different from the exact one. We will check a posteriori that these boundary conditions are indeed adapted to outgoing waves without spurious reflections. Moreover, we will demonstrate that the solution becomes independent of $R_2$, converging to what we will call the theoretical value of an infinite domain when $R_2\gg\rlight$.

\subsection{Divergencelessness constraint on $\textbf{\textit{D}}$ and $\textbf{\textit{B}}$}

In the special case of vacuum solutions, the electromagnetic fields $\textbf{\textit{D}}$ and $\textbf{\textit{B}}$ are divergencelessness. This property can be applied exactly analytically by the expansions below
\begin{subequations}
 \begin{align}
   \label{eq:D_div0}
  \textbf{\textit{D}} & = \sum_{l=1}^\infty\sum_{m=-l}^l \rot [f^D_{lm}(r,t) \, \mathbf{\Phi}_{lm}] + g^D_{lm}(r,t) \, \mathbf{\Phi}_{lm} \\
  \label{eq:B_div0}
  \textbf{\textit{B}} & = \sum_{l=1}^\infty\sum_{m=-l}^l \rot [f^B_{lm}(r,t) \, \mathbf{\Phi}_{lm}] + g^B_{lm}(r,t) \, \mathbf{\Phi}_{lm} 
\end{align}
\end{subequations}
where $\{f^D_{lm}(r,t), g^D_{lm}(r,t)\}$ and $\{f^B_{lm}(r,t), g^B_{lm}(r,t)\}$ are the expansion coefficients of $\textbf{\textit{D}}$ and $\textbf{\textit{B}}$ respectively. Thus, to impose the divergencelessness constraint, we project the electromagnetic field on to the subspace subtended by the expansions in equations~(\ref{eq:D_div0})-(\ref{eq:B_div0}). Actually, because spectral methods for smooth problems are very accurate, the projection is not required at each time step. We perform it only when the divergence becomes larger than a threshold defined by the user. Typically, we used a threshold of $10^{-10}$, normalized to the maximum value of the associated field, $|\divg \textbf{\textit{D}}|_{\rm max}/|\textbf{\textit{D}}|_{\rm max}$. When the maximum value of the divergence becomes greater than this value, we reinforce the constraint, otherwise we do not touch the corresponding field.

\subsection{Time integration}

One of the strength of pseudo-spectral methods is that they replace a set of partial differential equations (PDE) by a larger set of ordinary differential equations (ODE) for the unknown collocation points or spectral coefficients. Schematically, it can be written  as
\begin{equation}
  \label{eq:ODE}
  \frac{d\textbf{\textit{u}}}{dt} = f(t,\textbf{\textit{u}})
\end{equation}
with appropriate initial and boundary conditions. $\textbf{\textit{u}}$ represents the vector of unknown functions either evaluated at the collocation points or the spectral coefficients. Many time marching algorithms exist, see the discussion in \cite{2012MNRAS.424..605P}. Here we decide to use a fourth order Adam-Bashforth scheme advancing the unknown functions~$\mathbf{u}$ as
\begin{equation}
  \label{eq:AB4}
  \textbf{\textit{u}}^{n+1} = \textbf{\textit{u}}^{n} + 
  \frac{\Delta t}{24} \, \left( 55 \, \textbf{\textit{f}}^{n} - 59 \, \textbf{\textit{f}}^{n-1} + 37 \, \textbf{\textit{f}}^{n-2} - 9 \, \textbf{\textit{f}}^{n-3} \right)
\end{equation}
where $\Delta t$ is the fixed time step subject to some stability restrictions, $\textbf{\textit{u}}^{n} = \textbf{\textit{u}}(n\,\Delta t)$ and $\textbf{\textit{f}}^{n} = \textbf{\textit{f}}(n\,\Delta t,\textbf{\textit{u}}(n\,\Delta t))$. The three first steps are initialized with a classical fourth order Runge-Kutta method. The first and second order Adam-Bashforth scheme are weakly unstable for oscillatory solutions. To avoid such problems, it was necessary to use at least a third order algorithm as was done in our previous work \citep{2012MNRAS.424..605P}. Here we decide to use an even more accurate scheme for better comparison with previous semi-analytical results.

\subsection{Initial conditions}

We start our simulations with a neutron star initially at rest in vacuum. We then slowly and smoothly increase its spin frequency in order to avoid the formation of sharp gradients. This would be especially true at time $t=0$ where there is no electric field outside the star but right on its surface. Taking an evolution of the spin frequency as
\begin{equation}
 \Omega(t) = 
\begin{cases}
\sin^2\left(\frac{t}{8}\right) \text{ for } t\leq4\,\upi \\ 
1  \text{ for } t\geq4\,\upi
\end{cases}
\end{equation}
therefore starting at a null value avoids the initial discontinuity in the electric field. The spin frequency as well as its first derivative are smooth at the initial time of the simulation~$t=0$. No gradient or sharp features are expected. We next switch to a discussion of the results.

\section{RESULTS}
\label{sec:Results}

For the remaining of the paper, we adopt the following normalization:
the magnetic moment of the star is equal to unity, as well as the stellar angular
velocity and the speed of light, $\Omega=c=1$, therefore $\rlight=1$.

\subsection{The Deutsch solution}

Before handling the general-relativistic rotator, we checked our algorithm against some well known analytical solutions. The starting point is the Deutsch vacuum field solution for a perpendicular rotator. We recall that the Poynting flux of the magneto-dipole losses for an orthogonal rotator is
\begin{equation}
  \label{eq:L_dip}
  L_{\rm dip} = \frac{8\,\upi}{3} \, \frac{\Omega^4\,B^2\,R^6}{\mu_0\,c^3}.
\end{equation}
This does not include the boundary conditions on the neutron star. Actually it is only valid for a point dipole. The exact flux for the Deutsch field is
\begin{equation}
  \label{eq:L_deutsch}
 L_{\rm deustch} = \frac{4}{5} \, \frac{45 - 3\,x^4 + 2\,x^6}{(1 + x^2)\,(36 - 3\,x^4 + x^6)} \, L_{\rm dip}
\end{equation}
where $x=R/\rlight$. The finite size effect of the neutron star has a maximal impact when $R\lesssim\rlight$. It corresponds to the maximum allowed speed of a neutron star where the equatorial linear speed reaches the speed of light, which is evidently unrealistic. We will use at most $R=0.5\rlight$ not as a physically realistic case because it would mean a neutron star rotating at a sub-millisecond period which is not observed, but as a special case to test our algorithm against large variations in~$x$.

We start the simulation with the non rotating perpendicular dipole field, $\chi=90\degr$ and $\Omega=0$, and zero electric field outside the star, except for the crust where we enforce the inner boundary condition with corotating electric field, see equation~(\ref{eq:CLimites}). Note however that due to our special profile of $\Omega(t)$, the electric field at the surface of the star is equal to zero. It will slowly increase to its maximal value reached at a normalized time $t=4\,\upi$.

\begin{figure}
  \centering
  \includegraphics[angle=-90,width=0.5\textwidth]{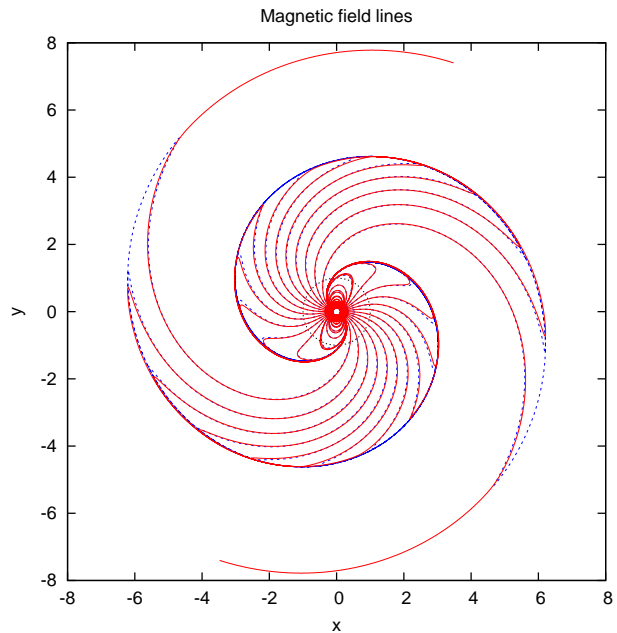}
  \caption{Magnetic field lines of the perpendicular Deutsch field in
    the equatorial plane for $\rlight/R=10$. The time-dependent simulation (solid red lines) is compared to the exact analytical solution (dashed blue lines). They are hardly distinguishable. The light-cylinder is shown as a dotted black circle of radius unity.}
  \label{fig:Deutsch_Ligne}
\end{figure}

\begin{figure}
  \centering
  \includegraphics[width=0.5\textwidth]{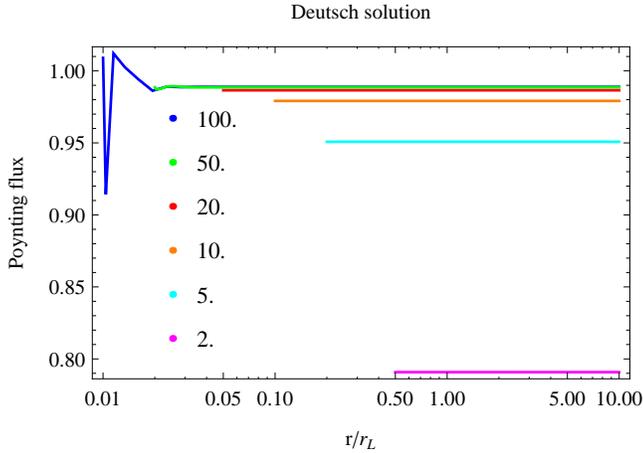}
  \caption{Normalized Poynting flux~$L/L_{\rm dip}$ across the sphere of radius~$r$, where $L$ is evaluated from eq.~(\ref{eq:Poynting}) and $L_{\rm dip}$ given by eq.~(\ref{eq:L_dip}). The inset legend corresponds to the ratio $\rlight/R=\{100,50,20,10,5,2\}$. The flux is perfectly constant in radius as is required by energy conservation. Its normalized value deviates strongly from unity only for the value $\rlight/R=2$. Note the logarithm scale in radius.}
  \label{fig:Deutsch_Poynting}
\end{figure}

We performed simulations with different spin frequencies of the neutron star corresponding to several ratio between stellar radius~$R$ and light cylinder radius~$\rlight$ such that $\rlight/R = \{100,50,20,10,5,2\}$. For the largest domain in radius with $r/\rlight\in[0.01,10]$ we found that a minimum resolution of $N_{\rm r} \times N_\vartheta \times N_\varphi = 257\times4\times8$ was necessary. We let the system evolve for about ten rotational period of the pulsar, $t_{\rm final} = 20\,\upi$. An example of final magnetic field line configuration in the equatorial plane is shown as a solid red curve in fig.~\ref{fig:Deutsch_Ligne} and compared to the analytical solution depicted as a dashed blue curve for $\rlight/R=10$. The numerical solution is very close to the exact solution everywhere, they are hardly distinguishable. Even by zooming into the light-cylinder region, both solutions agree very well with an overlap of magnetic field lines. The Poynting flux for the whole set of simulations is shown in fig.~\ref{fig:Deutsch_Poynting}. Note the logarithm scale in radius in order to better distinguish the different inner boundary locations of $R_1$. It is constant in radius, as expected from conservation of energy, and equal to the Deutsch value everywhere. Only for the case $\rlight/R=100$ is the numerical resolution on the edge of an acceptable accuracy close to the neutron star. This is cured by increasing the number of discretization points in the radial direction but we are limited by computational resources and prohibitive computation time. The system evolved to a steady state configuration and no significant reflections have to be reported at the outer boundary. The Poynting flux is a bit less than unity because it depends on the ratio~$R/\rlight$ and tends to Eq.~(\ref{eq:L_dip}) when $R/\rlight$ tends to zero as expected from equation~(\ref{eq:L_deutsch}).
\begin{figure}
  \centering
  \includegraphics[angle=-90,width=0.5\textwidth]{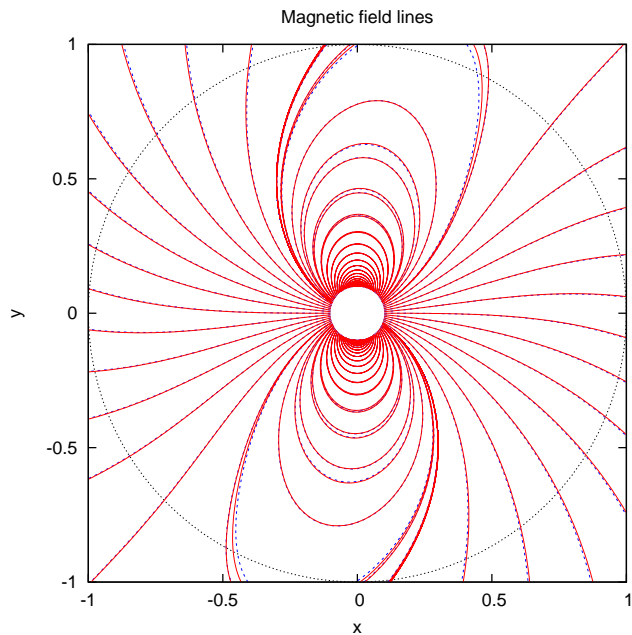}
  \caption{Zoom inside the light-cylinder of fig.~\ref{fig:Deutsch_Ligne} showing the good agreement between the analytical and numerical magnetic field structure in flat vacuum spacetime.}
  \label{fig:Deutsch_Ligne_Zoom}
\end{figure}
The table~\ref{tab:Flux_Poynting} compares the estimates of the Poynting flux from our simulations with the exact value given by Deutsch solution. The agreement is better than one percent whatever the size of the simulation box. 
\begin{table}
\begin{center}
\begin{tabular}{rrr}
\hline
$\rlight/R$ & Deutsch field & Simulations \\ 
\hline
\hline
  2 & 0.8010 & 0.7978 \\ 
  5 & 0.9615 & 0.9588 \\ 
 10 & 0.9901 & 0.9873 \\
 20 & 0.9975 & 0.9943 \\
 50 & 0.9996 & 0.9963 \\
100 & 0.9999 & 0.9968 \\
\hline
\end{tabular}
\end{center}
 \caption{Normalized Poynting flux~$L/L_{\rm dip}$ where $L$ is evaluated from eq.~(\ref{eq:Poynting}) and $L_{\rm dip}$ given by eq.~(\ref{eq:L_dip}) for the perpendicular rotator compared to the Deutsch solution. The agreement is excellent.}
 \label{tab:Flux_Poynting}
\end{table}
This paragraph showed that our pseudo-spectral code is mature and able to compute accurately vacuum electromagnetic fields in flat space-time. Boundary conditions have been implemented in a efficient way avoiding spurious reflections and artificial inner boundaries as usually required for finite difference/volume methods. Now it is time to look for the general-relativistic solution.

\subsection{General-relativistic solution}

This paragraph presents the new results extending the Deutsch solution to the general-relativistic dipole radiation in vacuum. We adopt the fixed background metric for a slowly rotating neutron star as described in section~\ref{sec:Modele}.

The same spin frequencies than those for the Deutsch solution are used, corresponding to $\rlight/R=\{100,50,20,10,5,2\}$. Although the largest values close to unity sound at the edge of the physically allowed or observed parameters, they serve as a good check of the algorithm because frame dragging effect are very significant for those values. The compactness, typical of a neutron star, is set to~$\Xi^{-1}=R/\Rs=2$.

\begin{figure}
  \centering
  \includegraphics[angle=-90,width=0.5\textwidth]{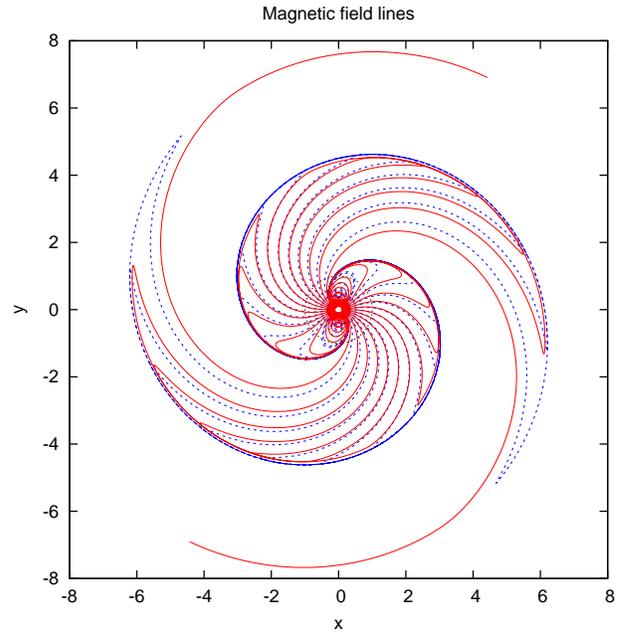}
  \caption{Magnetic field lines of the perpendicular general-relativistic field in the equatorial plane for $\rlight/R=10$. The time-dependent simulation, in solid red line, is compared to the exact analytical Deutsch solution, in dashed blue lines. The light-cylinder is shown as a dotted black circle of radius unity.}
  \label{fig:GR_Ligne}
\end{figure}

\begin{figure}
  \centering
  \includegraphics[angle=-90,width=0.5\textwidth]{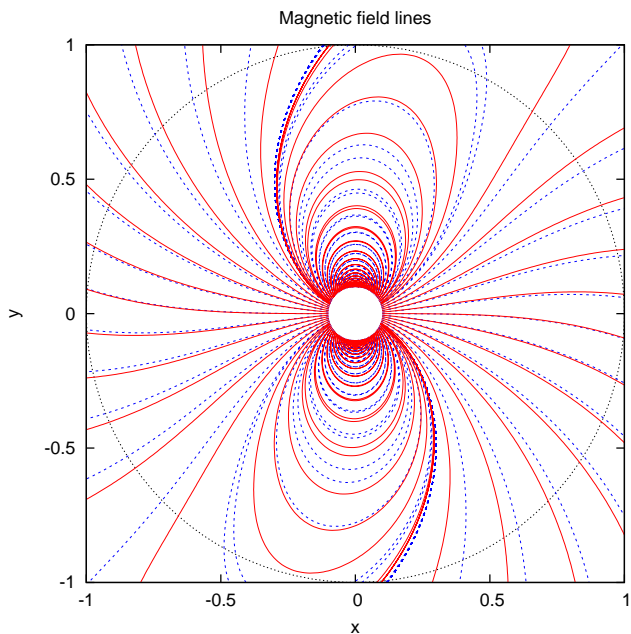}
  \caption{Zoom inside the light-cylinder of fig.~\ref{fig:GR_Ligne} showing the discrepancies between the Deutsch magnetic field structure and the general-relativistic structure.}
  \label{fig:GR_Ligne_Zoom}
\end{figure}

\begin{figure}
  \centering
  \includegraphics[width=0.5\textwidth]{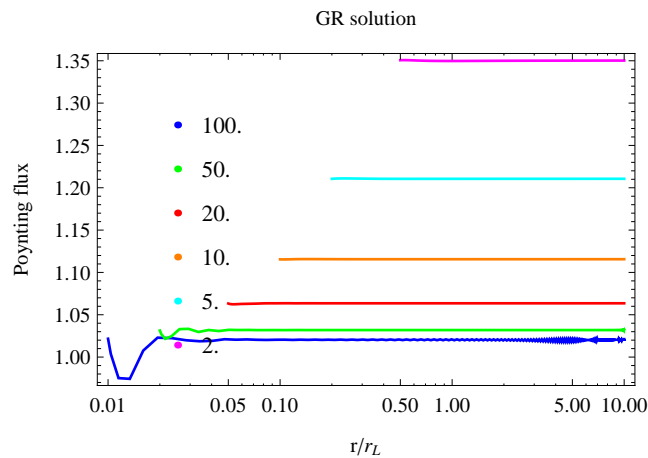}
  \caption{Normalized Poynting flux~$L/L_{\rm dip}$ across the sphere of radius~$r$  where $L$ is evaluated from eq.~(\ref{eq:Poynting}) and  $L_{\rm dip}$ given by eq.~(\ref{eq:L_dip}). The computed flux is constant as expected. The solution settled down to a stationary state. The inset legend corresponds to the ratio $\rlight/R = \{100,50,20,10,5,2\}$. Note the logarithm scale in radius.}
  \label{fig:GR_Poynting}
\end{figure}

First, the magnetic field line structure for the perpendicular rotator with $\rlight/R=10$ is shown in figure~\ref{fig:GR_Ligne}, to easily compare with figure~\ref{fig:Deutsch_Ligne}. The red solid lines depicts the general-relativistic solution obtained from our simulations. It is overlapped with the exact analytical Deutsch solution in dashed blue lines. We clearly recognize a deviation between both solutions. Nevertheless, the overall spiral structure of the field is preserved. Indeed, far from the neutron star, curved space-time effects become negligible and the solution relaxes to the usual Deutsch solution. The main difference resides in an azimuthal phase shift between both cases. The general-relativistic field lags slightly the flat space-time field according to figure~\ref{fig:GR_Ligne}. With no surprise, the strongest discrepancies arise close to the neutron star surface, where both curvature and frame dragging are important. Figure~\ref{fig:GR_Ligne_Zoom} indeed shows the region inside the light-cylinder. Strongest deviations from the flat spacetime field occur on magnetic field lines having their footpoints anchored in regions most distant from the magnetic poles. This could strongly impact on the radio pulse profiles and polarization properties around the polar caps. These observational implications give insight into the polar caps but are left for future investigations. 
Second, the Poynting flux as seen by an observer at infinity is plotted in figure~\ref{fig:GR_Poynting} for the set of ratio $\rlight/R = \{100,50,20,10,5,2\}$. It is constant with radius because measured by a distant observer, see the definition in \cite{2013MNRAS.433..986P}. It emphasizes again conservation of energy. Moreover here also the system evolved to a steady state configuration without reflection at the outer boundary. Our characteristics compatibility method used in flat space-time does also give good results in a curved space-time, when the outer boundary is kept far from the light cylinder, justifying its numerical use. For the ratio $\rlight/R = \{100, 50\}$, the computational resources are very demanding because the field is approximately increasing like $r^{-3}$ and includes a complicated radial dependence on $r$ with a logarithm function. The results obtained for the Poynting flux in these cases are less accurate as seen in figure~\ref{fig:GR_Poynting}.
Finally, the table~\ref{tab:Flux_Poynting_GR} summarizes the main result of the simulations, namely the Poynting flux as measured by a distant observer. A comparison between Deutsch solution, the semi-analytical computation in \cite{2013MNRAS.433..986P} and the present work demonstrates that for very slowly rotating neutron stars, i.e. $\rlight/R\gtrsim10$, results of semi-analytical computations and numerical time-dependent simulations are in agreement to each other. For fast rotating neutron stars, i.e. $\rlight/R\lesssim10$, deviations between semi-analytical and time-dependent simulations become significant. As was already pointed out in \cite{2013MNRAS.433..986P}, the convergence of the Chebyshev coefficients was observed to be poor, therefore we would expect the current values reported in the rightmost column of the table~\ref{tab:Flux_Poynting_GR} to be the most accurate ones. Moreover, our time-dependent simulations include many multipole electric and magnetic moments which was not the case for the semi-analytical results.
\begin{table}
\begin{center}
\begin{tabular}{rrrr}
\hline
$\rlight/R$ & Deutsch field & Semi-analytical & Simulations \\ 
\hline
\hline
  2 & 0.8010 & 1.4201 & 1.3502 \\ 
  5 & 0.9615 & 1.2541 & 1.2106 \\ 
 10 & 0.9901 & 1.1344 & 1.1156 \\
 20 & 0.9975 & 1.0792 & 1.0635 \\
 50 & 0.9996 & 1.0431 & 1.0319 \\
100 & 0.9999 & 1.0328 & 1.0211 \\
\hline
\end{tabular}
\end{center}
 \caption{Normalized Poynting flux where $L$ is evaluated from eq.~(\ref{eq:Poynting}) and  $L_{\rm dip}$ given by eq.~(\ref{eq:L_dip}), for the general-relativistic perpendicular rotator compared to the expectation from the Deutsch solution and our previous semi-analytical work. Neither gravitational redshift nor magnetic field amplification are taken into account here. This should emphasize the effect of frame-dragging only.}
 \label{tab:Flux_Poynting_GR}
\end{table}

\subsection{Influence of the location of the outer boundary}

\begin{figure}
 \centering
 \includegraphics[width=0.5\textwidth]{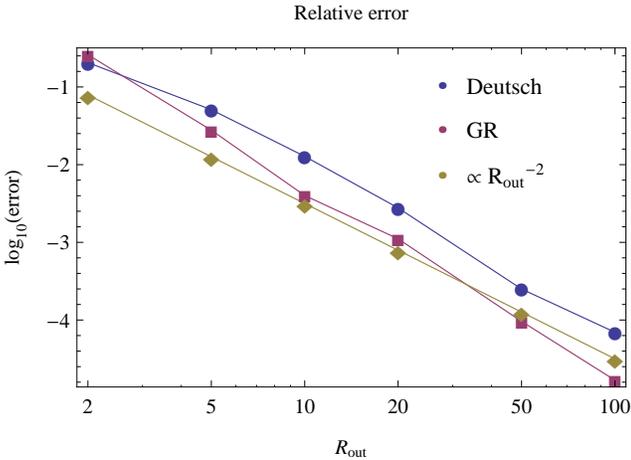}
 \caption{Evolution of the relative error between the exact (Deutsch) or expected (GR) Poynting flux and the flux obtained from the simulations.}
 \label{fig:ErreurRout}
\end{figure}

Imposing exact outgoing wave boundary conditions on a sphere of finite radius is a tedious work. Indeed \cite{2004JCoPh.197..186N} showed that the Sommerfeld radiation condition is only valid for the monopole field. For dipolar or even multipolar structures, restricting the infinite domain to a sphere of radius~$R_{\rm out}$ will lead to some deviation from a perfect outgoing wave. To elucidate the influence of the location of this outer sphere, we looked at the error of the Poynting flux with respect to the location of $R_{\rm out}$ defined by
\begin{equation}
\label{eq:erreur_rout} 
\epsilon = \left|\frac{L_{\rm ana} - L_{\rm num}}{L_{\rm ana}}\right|
\end{equation}
where $L_{\rm ana}$ and $L_{\rm num}$ are the analytical and numerical Poynting fluxes respectively.
The results are summarized in table~\ref{tab:Flux_Poynting_R_out} for the flat and curved spacetime, choosing a radius of the neutron star equal to $\rlight/R=2$ and $R_{\rm out}/\rlight = \{2,5,10,20,50,100\}$. For the Deutsch field we know the exact solution, see table~\ref{tab:Flux_Poynting}. We therefore plot the relative error between the truncated solution and the exact analytical solution against $R_{\rm out}$. The results are shown in figure~\ref{fig:ErreurRout}. We note that the error decreases like $R_{\rm out}^{-2}$. For the GR solution we extrapolate to the exact solution using this scaling (we found a value close to~$L_{\rm ana}\approx1.34478$) and show the corresponding relative error on the same plot. For $R_{\rm out}<10\,\rlight$, the precision is less than 1\% but for $R_{\rm out}>10\,\rlight$ reasonable good accuracy is obtained. The situation obviously improves when the sphere is translated more and more outwards but at the expense of higher resolution and longer cpu time simulations. In the previous results we kept $R_{\rm out}=10\,\rlight$ which is a good compromise between accuracy and computational time.

\begin{table}
\begin{center}
\begin{tabular}{rrr}
\hline
$R_{\rm out}/\rlight$ & Deutsch & GR \\ 
\hline
\hline
  2 & 0.6361 & 0.9951 \\ 
  5 & 0.8419 & 1.3821 \\ 
 10 & 0.7909 & 1.3502 \\
 20 & 0.8033 & 1.3463 \\
 50 & 0.8012 & 1.3449 \\
100 & 0.8011 & 1.3448 \\
\hline
\end{tabular}
\end{center}
 \caption{Normalized Poynting flux $L/L_{\rm dip}$, obtained from the simulations for different locations of the outer sphere~$R_{\rm out}$.}
 \label{tab:Flux_Poynting_R_out}
\end{table}

\section{DISCUSSION }
\label{sec:Discussion}

It is known that a neutron star cannot be surrounded by a strict vacuum outside its surface. Indeed, the electric field induced by the rotation of the magnetic field generates huge Lorentz forces able to extract particles from the crust and therefore filling the magnetosphere. 
Therefore the picture we present here can only be a starting point for more accurate solutions to the whole pulsar magnetosphere. Particles in this magnetosphere will produce a current disturbing the vacuum field, especially close to the light cylinder and outside it. A more realistic model would then account for such currents. Precisely how these charges flow is very model dependent. The simplest assumption would be to assume a force-free magnetosphere where the plasma dynamics is completely dominated by the electromagnetic field. The resulting force-free magnetosphere has been investigated by many authors like for instance in the aligned case by \cite{1999ApJ...511..351C} and the general oblique rotator by \cite{2006ApJ...648L..51S, 2009A&A...496..495K, 2012MNRAS.420.2793K, 2012MNRAS.424..605P}. Efforts towards a model leading to observational consequences include resistive solutions as done by \cite{2012ApJ...746...60L}. Energy should be dissipated into high-energy radiation. Some other used an MHD approach or PIC simulations. To date it is not clear whether the magnetosphere is partially or fully filled with plasma. Nevertheless
our present study clarifies the background structure of the electromagnetic field before perturbations imposed by the external current. The crucial question is about the particle density number in the magnetosphere. If too low, the picture presented in this paper seems to be a good zero order approximation of the electromagnetic field. If too high, the MHD or force-free solution should be investigated.

Thus in the optimistic case, the vacuum description of the electromagnetic field structure near the neutron star surface could be applied to model the polarization properties of the pulsed emission by curvature radiation emanating from the polar caps. Indeed, in the rotating vector model the polarization vector lies in a plane tangent to the curvature of the field line. Simple calculations did not include general-relativistic effects. This would be possible with the present results. Close to the neutron star, the magnetic field is not significantly distorted by the plasma current. It could be used to good accuracy for such a computation.
Moreover, further away from the neutron star, in the striped wind model, the location of the stripe should be slightly shifted compared to flat space-time due to the lag between the Deutsch solution and the one shown in the previous section. This induces an additional time lag between radio and gamma-ray pulses. We plan to quantify such delays in the near future, following the estimate done in \cite{2011MNRAS.412.1870P} for flat spacetime.

The next step would be to include the force-free approximation into this general-relativistic study. As mentioned by \cite{1990SvAL...16..286B}, these effects are crucial for the magnetic field aligned electric field and thus for particle creation, acceleration and radiation. Indeed, deviation from the corotation charge density leads to a parallel component of the electric field determined by the magnetic field geometry. Therefore space-time curvature and frame dragging effects are important for the electrodynamics of the gaps, see also \cite{1992MNRAS.255...61M}.

Last but not least, such computations will help to constrain light-curves of accreting millisecond pulsars by giving a precise and quantitative structure of the polar caps heated by the accreting plasma. One goal would be to get the mass to radius ratio $M/R$ and constrain the equation of state of neutron star matter \citep{2008ApJ...689..407B}.

\section{CONCLUSION}
\label{sec:Conclusion}

We solved the three-dimensional time-dependent Maxwell equations in spherical geometry and in the background space-time of a slowly rotating neutron star. We followed the 3+1~formalism according to an expansion of the unknown electromagnetic field on to vector spherical harmonics. We computed the corresponding spin-down luminosities and got results similar but slightly different from the semi-analytical analysis exposed in \cite{2013MNRAS.433..986P}. The general trend is an increase in the Poynting flux due to curvature and frame-dragging effects. In this way we improved the often quoted estimate of the magnetic dipole losses in flat space-time to a curved space-time suitable for a rotating neutron star.

The pulsar force-free magnetosphere could benefit from the same analysis. Numerical time-dependent simulations in the general-relativistic framework as done above but including the force-free current density allows a better quantitative description of the regions close to the neutron star surface. This is especially true at the polar caps, location supposed to host the coherent radio emission.
Such simulations will help to constrain the inner magnetosphere of radio pulsar and sharpens our understanding of their emission properties.

\section*{Acknowledgements}

I am grateful to the referee for helpful comments. This work has been supported in part by the French National Research Agency (ANR) through the grant No. ANR-13-JS05-0003-01 (project EMPERE).

\label{lastpage}

\end{document}